\documentstyle[aps,preprint,eqsecnum]{revtex}

\def\beq{\begin{equation}}
\def\eeq{\end{equation}}

\def\beqa{\begin{eqnarray}}
\def\eeqa{\end{eqnarray}}

\def\eq#1{(\ref{#1})}
\def\Eq#1{Eq.~\eq{#1}}
\def\Eqs#1{Eqs.~\eq{#1}}

\def\Ref#1{Ref.~\cite{#1}}

\def\hepph#1{hep-ph/#1}

\def\jref#1#2#3#4{{\it #1} {\bf #2}, #3 (#4)}

\def\NPB#1#2#3{\jref{Nucl.\ Phys.}{B#1}{#2}{#3}}

\def\PLB#1#2#3{\jref{Phys.\ Lett.}{#1B}{#2}{#3}}

\def\PRD#1#2#3{\jref{Phys.\ Rev.}{D#1}{#2}{#3}}

\def\PRL#1#2#3{\jref{Phys.\ Rev.\ Lett.}{#1}{#2}{#3}}

\def\etc{{\em etc\/}}
\def\ie{{\em i.e\/}.}
\def\eg{{\em e.g\/}.}

\def\to{\mathop{\rightarrow}}

\def\myint{\int\mkern-5mu}
\def\sfrac#1#2{{\textstyle\frac{#1}{#2}}}  

\def\Dsl{\hbox{\kern.1em/\kern-.7000em$D$}} 

\def\scr#1{{\cal #1}}

\def\mybar#1{\kern 0.8pt\overline{\kern -0.8pt#1\kern -0.8pt}\kern 0.8pt}
\def\sla#1{\raise.15ex\hbox{$/$}\kern-.57em #1}
\def\Sla#1{\kern.15em\raise.15ex\hbox{$/$}\kern-.72em #1}

\def\roughly#1{\mathrel{\raise.3ex\hbox{$#1$\kern-.75em%
    \lower1ex\hbox{$\sim$}}}}
\def\lsim{\roughly<}
\def\gsim{\roughly>}

\def\tr{\mathop{\rm tr}}

\def\avg#1{\langle #1 \rangle}

\def\al{\alpha}

\def\ga{\gamma}
\def\Ga{\Gamma}
\def\de{\delta}
\def\De{\Delta}

\def\ka{\kappa}
\def\la{\lambda}

\def\Si{\Sigma}
\def\th{\theta}

\def\Kahler{K\" ahler}

\def\hc{{\rm h.c.}}

\def\eV{{\rm \ eV}}

\def\MeV{{\rm \ MeV}}
\def\GeV{{\rm \ GeV}}
\def\TeV{{\rm \ TeV}}

\def\sec{{\rm \ sec}}

\input epsf.sty
\def\caption#1{{\small
        \centerline{\vbox{\baselineskip=12pt
        \vskip.15in\hsize=5.0in\noindent{#1}\vskip.1in }}}}

\begin{document}
\tighten
\preprint{\vbox{
\hbox{hep-th/9806398}}}

\title{Dynamical determination of the unification scale\\
by gauge-mediated supersymmetry breaking}

\author{Z. Chacko,%
\footnote{E-mail: {\tt zc8@umail.umd.edu}}
\ 
Markus A. Luty,%
\footnote{Sloan Fellow.
E-mail: {\tt mluty@physics.umd.edu}}
\ 
Eduardo Ponton%
\footnote{E-mail: {\tt eponton@wam.umd.edu}}
\medskip}

\address{Department of Physics\\
University of Maryland\\
College Park, Maryland 20742\medskip}

\date{June, 1998}

\maketitle

\begin{abstract}
\noindent
We propose a mechanism for generating the grand-unification
(GUT) scale dynamically from the Planck scale.
The idea is that the GUT scale is fixed by the vacuum expectation
value of a `GUT modulus' field whose potential is exactly
flat in the supersymmetric limit.
If supersymmetry is broken by gauge mediation, a potential for the
GUT modulus is generated at 2 loops, and slopes away from the origin
for a wide range of parameters.
This potential is stabilized by Planck-suppressed operators in the
\Kahler\ potential, and the GUT scale is fixed to be of order
$M_* / (4\pi^2)$
(where $M_* \sim 10^{18}\GeV$ is the reduced Planck scale)
independently of the supersymmetry breaking scale.
The cosmology of this scenario is acceptable if there is an
epoch of inflation with reheat temperature
small compared to the supersymmetry-breaking scale.
We construct a realistic GUT that realizes these ideas.
The model is based on the gauge group $SU(6)$, and solves the
doublet-triplet splitting problem by a sliding singlet mechanism.
The GUT sector contains no dimensionful couplings or tuned parameters,
and all mass scales other than the Planck scale are generated dynamically.
This model can be viewed as a realistic implementation of the inverted
hierarchy mechanism.
\end{abstract}

\pacs{?}

\section{INTRODUCTION}
\noindent
The study of supersymmetric grand unified theories (SUSY GUT's) has
received renewed impetus in recent years because of the striking
agreement between the observed values of the gauge couplings and
the predictions of the simplest supersymmetric unified theories
\cite{unity}.
Also, the recent revival of models of gauge-mediated SUSY
breaking has opened up new possibilities for the realization of SUSY
in nature \cite{oldGMSB,newGMSB,newnewGMSB}.
(For a review, see \Ref{GMSBrev}.)
The condition that gauge-mediated models can
be embedded in a GUT is often imposed, with the motivation that one does
not want to give up the successful prediction of gauge coupling
unification.
In this paper, we explore the possibility that there is a deeper
connection between grand unification and gauge-mediated SUSY
breaking.
We construct a model in which the dynamics that breaks SUSY is
also responsible for fixing the GUT scale via an `inverted hierarchy'
mechanism \cite{invert}.
The mechanism is very robust, and the specific model we construct to
implement it is quite simple.

The idea is that
the GUT scale is determined by minimizing the potential for
an almost-flat `GUT modulus' field $\Phi$ whose vacuum expectation value 
(VEV) determines the GUT scale.
It is assumed that SUSY is broken by the VEV of a singlet field $X$
below the GUT scale:
$F \equiv \avg{F_X} \ll \avg{\Phi}^2$.
The field $X$ has trilinear couplings to charged `messenger' fields,
and loop corrections involving gauge bosons give rise to supersymmetry
breaking terms in the observable sector.
In such models, SUSY breaking is communicated to $\Phi$ via 2-loop graphs,
which give $\Phi$ a logarithmic potential that
slopes away from the origin for a wide range of parameters:
\beq
V_{\rm GMSB}(\Phi) \sim -\frac{F^2}{(4\pi^2)^2} \ln^2 \Phi.
\eeq
If this were the only contribution to the potential, $\Phi$ would
run away to infinity.
However, in the context of supergravity (or string theory),
we expect the effective field
theory below the Planck scale to contain operators such as
\beq
\scr{L}_{\rm eff} \sim \pm \myint d^4\th\, \frac{1}{M_*^2}
X^\dagger X \Phi^\dagger \Phi
= \pm \frac{F^2}{M_*^2} \Phi^\dagger \Phi + \cdots,
\eeq
where
$M_* = M_{\rm Planck} / \sqrt{8\pi}
\simeq 2.4 \times 10^{18}\GeV$ is the reduced Planck mass.
This stabilizes the potential for $\Phi$ if the sign is negative, and gives
\beq
\avg{\Phi} \sim \frac{1}{4\pi^2} M_*.
\eeq
This relation is very robust:
it is independent of the SUSY breaking scale $F$, and
gives the right magnitude for the GUT scale as long as all
dimensionless couplings are order one.

The mass of $\Phi$ is
\beq
\label{mM}
m_\Phi \sim \frac{F}{M_*},
\eeq
so $m_\Phi$ is small compared to the weak scale precisely when
gauge-mediated contributions to SUSY breaking dominate
over the supergravity mediated contribution.
The interactions of $\Phi$ at low energies are given by
higher-dimension operators suppressed by powers of
$\avg{\Phi} \sim 10^{16}\GeV$.
This explains why the $\Phi$ particles cannot be seen directly in
laboratory experiments.

We construct a realistic GUT model that realizes these ideas.
The model is based on gauge group $SU(6)$, with the doublet-triplet
splitting problem solved using a version of the sliding singlet mechanism
\cite{slide} suggested by Barr \cite{Barr}.
The Higgs sector consists of
${\bf 35} \oplus {\bf 35} \oplus 4 \times ({\bf 6} \oplus \bar{\bf 6})$
plus singlets.
To incorporate 3 generations of matter, one needs
$3 \times ({\bf 15} \oplus \bar{\bf 6} \oplus \bar{\bf 6})$
and additional ${\bf 15} \oplus \bar{\bf 15}$ Higgs fields.
The model is therefore easily perturbative up to the Planck scale.
The Higgs superpotential contains only dimensionless couplings, and all
mass scales arise from the VEV of the field $\Phi$,
which is fixed dynamically by the mechanism above.
We do not specifically address the `$\mu$ problem,' but this can be
solved by simply adding a $\mu$ term to the model, or by one of the
mechanisms previously proposed in the literature \cite{muGMSB}.

Models with the mechanism described in this paper
suffer from a version of the well-known `Polonyi'
or `moduli' problem \cite{Polonyi}.
Because the potential for $\Phi$ is very flat, and the interaction
of $\Phi$ with other light fields is very weak at low energies,
it is not easy to understand why $\Phi$ is close to its minimum
in the early universe.
Coherent oscillations of $\Phi$ about the minimum can
dominate the energy density of the
universe, giving rise to an early matter-dominated era.
Nucleosynthesis is not possible during the $\Phi$-dominated era
\cite{nonuke},
and the eventual decay of $\Phi$ does not reheat the
universe sufficiently to allow nucleosynthesis.
These problems can be avoided by assuming an epoch of inflation
with low vacuum energy \cite{RT}.
If the vacuum energy during inflation and reheating is sufficiently
small, $\Phi$ is underdamped and relaxes to its minimum on a time scale
$1/H$, where $H$ is the expansion rate during inflation.
It is important to take into account the fact that
the minimum of the $\Phi$ potential is shifted from its
vacuum value due to supersymmetry breaking effects in the early
universe \cite{DRT}.
Nonetheless, inflation suppresses the energy density in
$\Phi$ oscillations and allows for a realistic cosmology.
As emphasized in \Ref{RT}, the density fluctuations that are the
seeds for structure formation can arise at a higher scale,
and mild constraints on the low-scale inflation ensure that the
fluctuations are not destroyed.
This model is therefore compatible with the presently-favored
scenario of structure formation seeded by inflation at high energy scales.

Of course, the present proposal is not the only possible mechanism
for fixing the GUT scale dynamically.
The GUT scale can emerge via dimensional transmutation in
weakly-coupled models \cite{weakdimtrans} or strongly-coupled
models \cite{strongdimtrans}.
In these models, the value of $M_{\rm GUT}$ depends sensitively
on the values of dimensionless couplings.
It has also been proposed that the potential for a GUT modulus
such as we are suggesting can arise from supergravity (or string)
effects \cite{KP}.
Our mechanism differs from these proposals in that the ratio
$M_{\rm GUT} / M_*$ is a robust prediction that does not depend on
details of Planck-scale physics.

This paper is organized as follows.
In Section II, we describe the model, compute and
minimize the potential for the GUT
modulus field, and construct the low-energy effective lagrangian.
In Section III we discuss the cosmology of this model.
Section IV contains our conclusions.

\section{THE MECHANISM}
In this Section we describe in detail the mechanism for fixing the
GUT scale in the context of a specific model.
The main ideas are more general than the model we present,
and we will emphasize the features that are important for our mechanism.

\subsection{Higgs Sector}
The first requirement for a successful model of the type outlined in the
Introduction is a Higgs sector that breaks a GUT group down to
$SU(3) \times SU(2) \times U(1)$ at a scale given by the
VEV of a `GUT modulus' flat direction $\Phi$.
In order to explain the success of gauge coupling unification,
we demand that the model incorporate a natural solution to the
doublet-triplet splitting problem.
Since we are trying to explain the origin of mass scales,
we demand that the Higgs sector contain no dimensionful couplings.

We now describe a simple model that satisfies all these requirements.
The model has gauge group $SU(6)$ and a Higgs sector consisting
of the following charged fields:
\beq
\Si,\ \De \sim {\bf 35},
\quad
H_{1,2},\ h_{1,2} \sim {\bf 6},
\quad
\bar{H}_{1,2},\ \bar{h}_{1,2} \sim \bar{\bf 6}.
\eeq
In addition, there are 8 singlets $\Phi$, $S$, $T_{1,2}$, $U_{1,2}$, and
$\bar{U}_{1,2}$.
The superpotential for the Higgs sector is
\beqa
W_{\rm Higgs} &=& \sfrac{1}{2} S (\tr \Si^2 - \Phi^2)
+ \sfrac{1}{6} \tr\Si^3
\nonumber\\
&& \qquad +\, \sum_{J = 1}^2
T_J ( \bar{H}_J H_J - \sfrac{1}{2} \Phi^2 )
\nonumber\\
&& \qquad +\, \sum_{J = 1}^2 \left[
\bar{H}_J \Si h_J + U_J \bar{H}_J h_J
+ \bar{h}_J \Si H_J + \bar{U}_J \bar{h}_J H_J \right]
\nonumber\\
\label{WHiggs}
&& \qquad +\, \sfrac{1}{2} \Phi \tr\De^2
+ ( \bar{H}_1 \De H_2 - \bar{H}_2 \De H_1 ).
\eeqa
(We will discuss fermion masses below.)
This superpotential is invariant under the ${\bf Z}_2$ symmetry
\beq
\label{discrete}
H_1 \leftrightarrow H_2,
\quad
\bar{H}_1 \leftrightarrow \bar{H}_2,
\quad
h_1 \leftrightarrow h_2,
\quad
\bar{h}_1 \leftrightarrow \bar{h}_2,
\quad
U_1 \leftrightarrow U_2,
\quad
\bar{U}_1 \leftrightarrow \bar{U}_2,
\quad
\De \leftrightarrow -\De,
\eeq
with all other fields invariant.
Because the superpotential contains only dimension-3 terms,
it is also invariant under a $U(1)_R$ symmetry under which all
fields have charge $\frac{2}{3}$.
\Eq{WHiggs} is not the most general superpotential allowed by
symmetries (\eg\ a $S^3$ term is allowed).
However, we do not consider this to be problematic since perturbative
non-renormalization theorems and their
non-perturbative generalizations \cite{NPNR} allow the superpotential
to naturally be `non-generic.'

The freedom to rescale the fields has been used to set
all Yukawa couplings to 1.
The information about the strength of the superpotential couplings is
therefore contained in the normalization of the kinetic terms.

\newpage
This potential has supersymmetric minima with
\beq\label{vevs}
\avg{\Si} = \frac{\avg{\Phi}}{\sqrt{6}}
\pmatrix{{\bf 1}_3 & 0 \cr 0 & -{\bf 1}_3 \cr},
\quad
\avg{H_{1,2}} = \frac{1}{\sqrt{2}} \avg{\bar{H}_{1,2}}
= \pmatrix{0 \cr \vdots \cr 0 \cr \avg{\Phi} \cr},
\quad
\avg{U_{1,2}} = \avg{\bar{U}_{1,2}} = \frac{\avg{\Phi}}{\sqrt{6}},
\eeq
and all other VEV's vanishing.
This breaks $SU(6)$ down to $SU(3) \times SU(2) \times U(1)$.
In the supersymmetric limit, there is a flat direction along which $\Phi$
changes, and $\avg{\Phi}$ is undetermined at this stage.
We therefore refer to the field $\Phi$ as the `GUT modulus.'
We require $\avg{\Phi} \sim 10^{16}\GeV$ for a successful model.

The roles of the various terms in \Eq{WHiggs} are not hard to understand.
The terms of the form $S (\Phi^2 - \tr \Si^2)$ and
$T ( \bar{H} H - \sfrac{1}{2} \Phi^2 )$ force
$\avg{\Si} \sim \avg{H} \sim \avg{\bar{H}} \sim \avg{\Phi}$.
The terms $\bar{H} \Si h + U \bar{H} h$
and $\bar{h} \Si H + \bar{U} \bar{h} H$ force
\beq
\avg{\Si + U_{1,2} {\bf 1}_6} =
\avg{\Si + \bar{U}_{1,2} {\bf 1}_6} =
\frac{2 \avg{\Phi}}{\sqrt{6}}\pmatrix{{\bf 1}_3 & 0 \cr 0 & 0 \cr},
\eeq
which gives mass terms to the Higgs triplets while leaving the doublets
massless.
This is a version of the sliding singlet mechanism \cite{slide}
due to Barr \cite{Barr}.
Finally, the terms involving $\De$ are needed to give mass to two
doublet components of the fields $H_{1,2}$, $\bar{H}_{1,2}$.
(In the model of \Ref{Barr}, one doublet gets a mass from a
higher-dimension operator, potentially upsetting gauge coupling
unification.)

A nontrivial feature of the superpotential \Eq{WHiggs} is that the
VEV of $\Phi$ is not fixed in the supersymmetric limit.
This is ensured by coupling $\Phi$ only
to fields that have zero VEV in the desired vacuum.
This requirement is nontrivial for the term $S (\Phi^2 - \tr \Si^2)$,
since the equation of motion for the diagonal components of $\avg{\Si}$
is a quadratic equation, and $\avg{S}$ is proportional to the sum
of the roots.
We have checked explicitly that the only massless fields correspond
to the flat direction and a pair of doublets
\beq
h^\al \equiv \frac{1}{\sqrt{2}} (h_1 - h_2)^\al,
\quad
\bar{h}_\al \equiv \frac{1}{\sqrt{2}} (\bar{h}_1 - \bar{h}_2)_\al,
\quad
\al = 4,5.
\eeq
These are the features we need for our mechanism
to fix the GUT scale dynamically.

Note that the $U(1)_R$ symmetry is spontaneously broken, and so the
model as written
has an $R$ axion with a decay constant of order $M_{\rm GUT}$.
However, the $U(1)_R$ symmetry may be explicitly broken by dimension-5
terms in the effective \Kahler\ potential that couple the GUT modulus
to the field $X$ that is responsible for SUSY breaking:
\beq
\de \scr{L}_{\rm eff} = \myint d^4\th
\left[
\frac{a_1}{M_*} X^\dagger \Phi^2
+ \frac{a_2}{M_*} X^\dagger X \Phi
+ \hc \right].
\eeq
The term proportional to $a_1$ gives the $R$ axion a mass of order
$m_\Phi \sim F / M_*$ if $a_1 \sim 1$.
The term proportional to $a_2$ gives a large shift to the vacuum found
above unless $a_2 \ll g^2 / (4\pi^2)$.
It is natural to have $a_2 \ll a_1$ because the terms have different
charges under global (discrete) symmetries.%
\footnote{We have checked that there is a discrete $R$ symmetry
that forbids $a_2$
under which the superpotential \Eq{WHiggs}
(and \Eqs{Wferm} and \eq{Wmess} below) is invariant.}
Therefore, the $R$ axion mass can be anywhere below $m_\Phi$
depending on the pattern of breaking of global symmetries
at the Planck scale.

\subsection{Fermion Masses}
Although we view the model above mainly as an illustration, it is
worth noting
that fermion masses can be incorporated without
significantly complicating the Higgs sector.
We add the fields
\beq
N_j \sim {\bf 15},
\quad
\bar{P}_{1j}, \bar{P}_{2j} \sim \bar{\bf 6},
\quad
Y \sim {\bf 15},
\quad
\bar{Y} \sim \bar{\bf 15},
\eeq
where $j = 1,2,3$ is a generation index.
(Note that there must be two $\bar{\bf 6}$'s for each ${\bf 15}$ to
satisfy $SU(6)$ anomaly cancellation.)
The additional superpotential terms required are
\beqa
W_{\rm fermion} &=& \la^{jk} N_j
(\bar{P}_{1k} \bar{H}_1 + \bar{P}_{2k} \bar{H}_2)
\nonumber\\
&&\qquad
+\, y_d^{jk} N_j (\bar{P}_{1k} \bar{h}_1 + \bar{P}_{2k} \bar{h}_2)
\nonumber\\
\label{Wferm}
&&\qquad
\label{Yuk}
+\, y_u^{jk} N_j N_k Y
+ \Phi Y \bar{Y} + \bar{Y} (H_1 h_1 - H_2 h_2).
\eeqa
The discrete symmetry \Eq{discrete} is extended to the additional fields
via
\beq
N \mapsto -iN,
\quad
\bar{P}_1 \mapsto i\bar{P}_2,
\quad
\bar{P}_2 \mapsto i\bar{P}_1,
\quad
Y \mapsto -Y,
\quad
\bar{Y} \mapsto -\bar{Y}.
\eeq
(The symmetry is therefore ${\bf Z}_4$
rather than ${\bf Z}_2$.)
We have scaled the fields $Y$ and $\bar{Y}$ to set some of the
superpotential couplings to 1.

The roles of the terms in \Eq{Yuk} are as follows.
The $N (\bar{P}_1 \bar{H}_1 + \bar{P}_2 \bar{H}_2)$ term gives a mass
of order $\avg{\Phi}$ to the unwanted components of the ${\bf 15}$ and
two ${\bf 6}$'s, leaving 3 generations of quarks and leptons massless.
The remaining terms give rise to the fermion Yukawa couplings.
The up-type Yukawa couplings arise when the massive fields
$Y$ and $\bar{Y}$ are integrated out.
It is not hard to see that adding these terms to the superpotential
preserves the vacuum described above.

Like most SUSY GUT's, the present model has a potential problem with
proton decay from the effective dimension-5 operator
\beq
W_{\rm eff} \sim \frac{y_u^{jk} y_d^{\ell m}}{\avg{\Phi}}
Q_j Q_k Q_\ell L_m
\eeq
arising from exchange of heavy color triplet Higgs fields.

As written, this model embodies unsuccessful $SU(5)$ GUT relations
among fermion masses, namely the equality of lepton and down-type
quark masses (up to radiative corrections).
As is well-known, this actually works well for the third generation but
not for the first two generations.
This can be remedied by assuming that there are
higher-dimension operators such as
\beq
\de W_{\rm fermion} = \frac{c^{jk}}{M_*} N_j \Si
(\bar{P}_{1k} \bar{h}_1 - \bar{P}_{2k} \bar{h}_2),
\eeq
whose contribution to the Yukawa couplings can be important for the
light fermions.

\bigskip\medskip
\centerline{\epsfbox{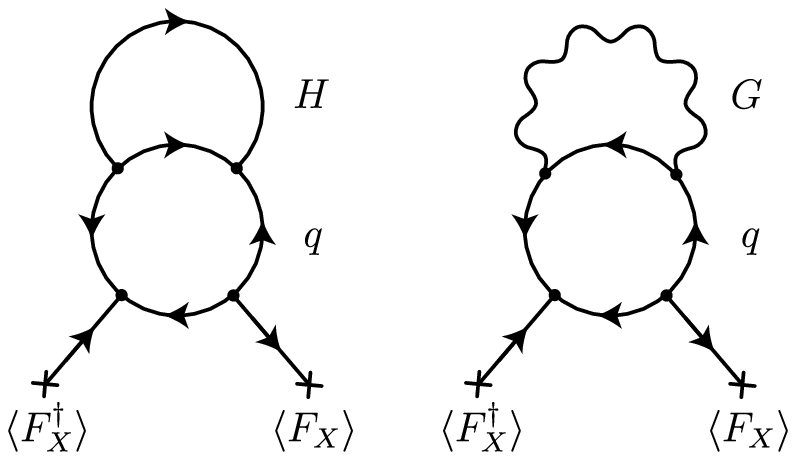}}
\smallskip
\caption{{\bf Fig.~1.}  Contributions to the effective potential of
$\Phi$ in gauge-mediated models.
Here $q$ denotes a messenger field, $H$ denotes a superheavy Higgs field,
and $G$ denotes a superheavy gauge boson field.
The dependence on $\Phi$ enters because the masses of $H$ and $G$ are
proportional to $\avg{\Phi}$.
The $H$ contribution drives $\Phi \to \infty$, while the $G$ contribution
drives $\Phi \to 0$.}
\bigskip

\subsection{Fixing the GUT Modulus}
In the class of models we are considering
the VEV of the GUT modulus field is undetermined in the SUSY limit,
but SUSY breaking will ultimately lift all flat directions and
determine the GUT scale.
If supersymmetry is broken in a hidden sector and communicated to
the observable sector by supergravity, the leading contribution to the
potential for $\Phi$ is expected to arise from effective \Kahler\ terms
of the form
\beq\label{Keff}
\scr{L}_{\rm eff} = \myint d^4\th\, \frac{c}{M_*^2}
X^\dagger X \Phi^\dagger \Phi + \scr{O}(1/M_*^3)
= \frac{c |F|^2}{M_*^2} \Phi^\dagger \Phi + \cdots,
\eeq
where $X$ is a field whose VEV breaks SUSY:
\beq
F \equiv \avg{F_X} \ll \avg{X}^2.
\eeq
The potential \Eq{Keff} drives $\Phi$ either to the origin or to
infinity depending on
the sign of $c$, so this does not give a mechanism to determine the GUT
scale.
If $\Phi$ is driven to infinity, new physics presumably enters
for $\avg{\Phi} \gsim M_*$, but it is not easy to see how this could
lead to $\avg{\Phi} \sim 10^{-2} M_*$.%
\footnote{\Eq{Keff} is expected to be the leading correction to the
\Kahler\ potential as long as
all field strengths are small compared to $M_*$ in models where the
only new physics (compactified dimensions, excited string modes,
\etc.) arise at or above the scale $M_*$.
If there are additional scales or light fields coupling the `hidden'
and observable sectors, then there may be a stable minimum for the
GUT modulus below $M_*$.}

However, we now describe a natural mechanism to fix the GUT modulus
at the correct value
in theories of gauge-mediated supersymmetry breaking.
In these models, the leading contribution to the potential for
$\Phi$ comes from the 2-loop graphs shown in Fig.~1.
We assume that $X$ is
coupled to charged messenger fields $q$, $\bar{q}$ via
\beq
W_{\rm mess} = \la X \bar{q} q.
\eeq
Using the techniques of \Ref{supercalc} this contribution to the effective
potential can be computed by solving 1-loop renormalization group
equations analytically continued into superspace.
(For earlier calculations, see \Ref{dircalc}.)
The result is
\beq\label{Mpot}
|\Phi| \frac{\partial V}{\partial |\Phi|}
\simeq \sum_q \frac{|\la|^2 |F|^2}{4\pi^2}
\biggl[ \ga_q(|\Phi|) - \ga_{q,{\rm eff}} (|\Phi|) \biggr]
\ln \frac{|\Phi|}{|\la X|},
\eeq
where the sum is over all components of the messenger field $q$, and
\beq
\ga_q(\mu) \equiv \mu \frac{d\ln Z_q}{d\mu}
\eeq
is the messenger anomalous dimension.
Here, $\ga_q$ ($\ga_{q,{\rm eff}}$) refers to the anomalous dimension
in the theory above (below) the GUT threshold evaluated at
$\mu = |\Phi|$.
In \Eq{Mpot} we have not summed logs of $|\la X| / |\Phi|$, and we have
identified the GUT threshold with $\Phi$ (rather than
Yukawa couplings times $\Phi$).
These effects can be easily included, but they are not important at the
level of accuracy we are working.

Because the potential is proportional to the difference of messenger
anomalous dimensions above and below the scale $\Phi$, it can be viewed
as arising from fields that get massive at the scale $\Phi$.
This makes the sign easy to understand.
Gauge loops give a positive contribution to $\ga_q$, while matter loops
give a negative contribution.
The potential will therefore slope away from the origin provided that
the contribution from matter loops dominates, a condition that is
easily satisfied (see below).

For sufficiently large values of $\Phi$, the supergravity contribution
\Eq{Keff} becomes important.
We assume $c < 0$, which corresponds to a positive value for the
supergravity-induced mass for $Y$.
This stabilizes the runaway behavior of the potential and fixes
$\avg{\Phi}$.
Neglecting couplings and group theory factors, $\ga_q \sim 1/(4\pi^2)$,
and the minimum is
\beq
\avg{\Phi} \sim \frac{M_*}{4\pi^2} \left(
\ln \frac{|\avg{\Phi}|}{|\avg{X}|} \right)^{1/2}.
\eeq
Since $M_* / (4\pi^2) \simeq 6 \times 10^{16}\GeV$,
this nicely explains the `observed' value
$\avg{\Phi} \simeq 3 \times 10^{16}\GeV$.
It is remarkable that the result is independent of $F$, and depends only
logarithmically on $\avg{X}$.

For example, in our $SU(6)$ model, the simplest possibility is to take
\beq
q \sim {\bf 6},
\quad
\bar{q} \sim \bar{\bf 6},
\eeq
and couple the messengers to the adjoint Higgs field $\De$
as well as the singlet field $X$ responsible for SUSY breaking:
\beq\label{Wmess}
W_{\rm mess} = \ka \bar{q} \De q + \la X q \bar{q}.
\eeq
(Recall that $\avg{\De} = 0$, so this does not generate a superheavy
mass for the messengers.)
We therefore have
\beq
\ga_q = \frac{C_q}{4\pi^2}\, g^2
- \frac{C_q}{8\pi^2}\, |\ka|^2,
\eeq
where $C_q = (N^2 - 1) / (2N)$ is the messenger Casimir.
Below the scale $\Phi$, there are contributions to the $q$ anomalous
dimension from the unbroken
$SU(3) \times SU(2) \times U(1)$ gauge fields, and we obtain
\beq
\sum_q \biggl[ \ga_q(|\avg{\Phi}|) - \ga_{q,{\rm eff}}(|\avg{\Phi}|) \biggr]
= \frac{1}{4\pi^2} \left(
\frac{23}{2} g^2 - \frac{35}{4} |\ka|^2 \right).
\eeq
This is naturally negative for $\ka \sim 1$.
Assuming that the contribution proportional to $\ka$ dominates,
we obtain
\beq
|\avg{\Phi}| \simeq \frac{M_* |\la\ka|}{4\pi^2} \left(
\frac{35}{8 |c|}\, \ln \frac{|\avg{\Phi}|}{|\la\avg{X}|} \right)^{1/2}.
\eeq
We see that a realistic GUT scale does in fact emerge for reasonable
values for the couplings.

\subsection{Interactions of the GUT Modulus}
From \Eqs{Keff} and \eq{Mpot}, we see that the mass of the GUT modulus is
\beq
m_\Phi \sim \frac{F}{M_*},
\eeq
\ie\ the same magnitude as the SUSY-breaking masses
communicated by supergravity.
Because we want gauge-mediation to dominate SUSY breaking in the
observable sector, we have $m_\Phi \lsim 100\GeV$.
Also, the smallest possible value for $F$ is of order $(10\TeV)^2$,
which gives
\beq
10^{-1}\eV \lsim m_\Phi \lsim 100\GeV.
\eeq

The existence of such a light particle is not immediately ruled out
because it interacts with other light fields only through
higher-dimension operators suppressed by powers of $1/\avg{\Phi}$.
This follows from the double role of the field $\Phi$.
On the one hand, the VEV of $\Phi$ sets the scale for all the superheavy
masses.
On the other hand, if we write
\beq
\Phi = \avg{\Phi} + \Phi',
\eeq
$\Phi'$ can only appear in the
combination $\avg{\Phi} + \Phi'$ because of the invariance under simultaneous
shifts of $\avg{\Phi}$ and $\Phi'$ that keep $\Phi$ fixed.%
\footnote{The precise statement is that it is possible to choose the
fields in the low-energy effective theory to have this property.}
This means that we can determine the $\Phi$ dependence in the effective
theory from the dependence on the GUT threshold.
The leading dependence on $\Phi$ comes from the fact that the dimensionless
couplings in the low-energy theory depend logarithmically on $|\Phi|$ from
the renormalization group evolution from the GUT scale.
This gives rise to interactions such as
\beq
\scr{L}_{\rm eff} \sim \frac{g^2}{4\pi^2} \frac{\Phi'}{\avg{\Phi}}
F^{\mu\nu} F_{\mu\nu},
\eeq
where the factor $\Phi' / \avg{\Phi}$ arises from a difference of
anomalous dimensions, as in the calculation of the potential above.
The scalar component of the
GUT modulus can therefore decay to photons with rate
\beq\label{Mdecay}
\Ga(\phi_\Phi \to \ga\ga) \sim \frac{1}{4\pi} \left( \frac{\al}{\pi} \right)^2
\frac{m_\Phi^3}{\avg{\Phi}^2}
\sim \frac{1}{2 \times 10^{45}\sec}
\left( \frac{\sqrt{F}}{10\TeV} \right)^{6}.
\eeq
Decay widths to other modes (such as $e^+ e^-$ and
$\bar{\nu}\nu$) are comparable.
The fermion component of $\Phi$ is expected to be stable, since there
is generally no lighter fermion that it can decay into.

\section{COSMOLOGY}
The potential for the field $\Phi$ is very flat, and the interaction of
$\Phi$ with other light fields is very weak at energies below the
GUT scale.
This leads to a potential cosmological problem known as the
`Polonyi' or `moduli' problem \cite{Polonyi}.
The problem is to understand why $\Phi$ is close to the minimum
of its potential in the early universe.
$\Phi$ will begin to oscillate about the minimum of its potential
when the expansion rate becomes smaller than its mass.
The coherent oscillations scale like non-relativistic matter,
so $\Phi$ oscillations will dominate the
energy density unless $\Phi$ is very close to its minimum
in the early universe.
If this condition is not satisfied,
$\Phi$ oscillations dominate the universe,
and the eventual $\Phi$ decay
reheats the universe.
The reheat temperature is
\beq
T_{\Phi,{\rm RH}} \sim 
\frac{\sqrt{ \Ga_\Phi M_*}}{g_*^{1/4}(\Phi,{\rm RH})}
\sim (10 \eV) \left( \frac{\sqrt{F}}{10^{10}\GeV} \right)^3.
\eeq
This is far too low to allow nucleosynthesis after $\Phi$ decay, and
successful nucleosynthesis is impossible during the $\Phi$-dominated
era \cite{nonuke}.

The natural framework for solving this problem is low-scale
inflation \cite{RT}.
During the slow-roll phase of inflation, the $\Phi$ equation of
motion is
\beq
\ddot{\Phi} + 3 H_{\rm inf} \dot{\Phi}
+ \frac{\partial V}{\partial \Phi} = 0,
\eeq
where
$H_{\rm inf}$ is the expansion rate of the universe during inflation, and
we have neglected the $\Phi$ decay term.
If $H_{\rm inf} \lsim m_\Phi$, the equation for $\Phi$ is
underdamped or critically damped,
and $\Phi$ approaches the minimum of its potential
on a time scale $1 / (3 H_{\rm inf})$.
As long as inflation persists for several $e$-folds, $\Phi$ will
be driven close to its minimum.
In terms of microscopic parameters, the condition for this to occur is
\beq\label{udamp}
V_{\rm inf} \lsim F^2,
\eeq
where $V_{\rm inf}$ is the vacuum energy that drives inflation.

If $V_{\rm inf} \gg F^2$, then
$H_{\rm inf} \gg m_\Phi$ and the evolution of $\Phi$ is overdamped.
$\Phi$ therefore approaches its minimum on a time scale
$3 H / m_\Phi^2 \gg 1/H$.
We will see that this scenario is ruled out even if we accept the
enormous number of $e$-folds required for $\Phi$ to relax to its
minimum.

The effective potential
for $\Phi$ during inflation is not the same as the vacuum potential.
This is important because the flatness of the $\Phi$
potential is protected by supersymmetry, which is explicitly broken
by the finite energy density in the early universe \cite{DRT}.
In general, there will always be a contribution to the effective potential
for $\Phi$ of the form
\beq
\Phi \frac{\partial V_{\rm eff}}{\partial \Phi}
\propto \rho,
\eeq
where $\rho$ is the energy density.
It will be important for us to have an estimate of the constant
of proportionality during the slow-roll phase of inflation.
For this we need to know the couplings between the
inflaton field $I$ and the GUT modulus $\Phi$.
The most conservative possible
assumption is that $\Phi$ couples to the inflaton
as strongly as to visible fields.
For example, if the inflaton couples to superheavy fields at the
GUT scale with dimensionless coupling constant $h$,
we obtain
\beq\label{VIpert}
\Phi \frac{\partial V_{\rm eff}}{\partial \Phi}
\sim \frac{h^2}{4\pi^2} V_{\rm inf}.
\eeq
We will use this formula to parameterize
the effect of the inflation energy on the $\Phi$ potential in
general models.
The inflaton may couple very weakly to $\Phi$,
and it is important to keep in mind that $h \ll 1$ is a
natural possibility.

Even though $\Phi$ is not at its vacuum value, it will track the
instantaneous minimum of its effective potential as long as the
potential is changing sufficiently slowly in time.
This will be the case as long as
\beq
\frac{\ddot{\rho}}{\rho} \ll m_\Phi^2.
\eeq
Since $\dot{\rho} \lsim H \rho$,
this is satisfied provided $H \ll m_\Phi$.
In order for this condition to be satisfied at the end of inflation,
we require
\beq\label{infbound}
V_{\rm inf} \ll F^2.
\eeq
With the suppression of $\Phi$ oscillations guaranteed by low-scale
inflation, the fact that $\Phi$ has a long lifetime and decays into
photons does not cause problems.

This scenario potentially suffers from a naturalness problem,
since terms in the effective theory of the form
\beq\label{badterm}
\de\scr{L} \sim \myint d^4\th\, \frac{1}{M_*^2}
X^\dagger X I^\dagger I
\eeq
naturally imply
$\partial^2 V / \partial I^2 \gsim {F^2}/{M_*^2}$,
which contradicts the slow-roll condition
$ \partial^2 V / \partial I^2 \ll V_{\rm inf} / M_*^2 \ll {F^2}/{M_*^2}$.
This may be avoided if the inflaton is a pseudo-Nambu-Goldstone
boson associated with the breaking of an approximate global symmetry.
The flatness of the inflaton potential is then unaffected by terms
such as \Eq{badterm} that do not violate the global symmetry.
Terms that do violate the symmetry may be naturally small.
Alternatively, terms such as \Eq{badterm}
may be naturally smaller than expected
on the basis of dimensional analysis if the inflaton sector and the
SUSY breaking sector are close to a limit where they decouple.

If the condition \Eq{infbound} on the inflation energy
is not satisfied, then the time-dependent
effective potential will cause $\Phi$ to oscillate with an amplitude
of order the difference between its value after inflation and its
vacuum value:
\beq\label{dPhi}
\De \Phi \sim h^2 \frac{V_{\rm inf}}{F^2}\, M_*,
\eeq
where we have assumed that the dimensionless couplings
$g$, $\la$, and $\ka$ (defined above) are of order 1.
The $\Phi$ energy density at the end of inflation is therefore
\beq
\left. \frac{\rho_\Phi}{\rho} \right|_{\rm inf}
\sim h^4 \frac{V_{\rm inf}}{F^2}.
\eeq
This need not dominate the present energy of the universe if
$h \ll 1$, \ie\ the inflaton is very weakly coupled to $\Phi$.
As an illustration, we consider the case $V_{\rm inf} \sim F^2$.
The field $\Phi$ then begins to oscillate immediately
after inflation, and we obtain
\beq
\left. \frac{\rho_\Phi}{\rho} \right|_{0}
\sim \frac{(g_*^{1/3} T)_{\rm RH}}{(g_*^{1/3} T)_{\rm EQ}}
\left. \frac{\rho_\Phi}{\rho} \right|_{\rm RH}
\sim 10^{10} h^4
\left( \frac{T_{\rm RH}}{100\GeV} \right),
\eeq
since the oscillations of both $\Phi$ and the inflaton scale
like non-relativistic matter during the reheating phase of
inflation.
We see that having a reheating temperature above the weak scale
requires $h \lsim 10^{-3}$.
Recall that $h \sim 1$ is the maximal possible coupling of the
inflaton to $\Phi$, so such a weak coupling is not unnatural.

In this scenario, the $\Phi$ oscillations are not necessarily
suppressed, and we must consider the consequences of the $\Phi$
decay into photons.
For $F \lsim 10^{8}\GeV$, the lifetime is longer than the present age
of the universe, and $\Phi$ decays are presently producing photons.
Assuming that the photon energy is $E_\ga \sim m_\Phi$,
the bound on the present energy in $\Phi$ oscillations is \cite{KT}
\beq
\frac{\rho_\Phi \Ga_\Phi}{4\pi H_0}
\lsim \frac{1\MeV}{1\ {\rm cm}^2\ {\rm sec}}.
\eeq
For the case $V_{\rm inf} \sim F^2$ considered above, this gives
the bound
\beq
h \lsim 10^{-2} \left( \frac{T_{\rm RH}}{100\GeV} \right)^{-1/4}
\left( \frac{\sqrt{F}}{10^7\GeV} \right)^{-3/2}.
\eeq

We have not separately discussed the interactions of the $R$ axion
in our model.
If its mass is of order $m_\Phi$, then the considerations above apply
equally to the real and complex components of $\Phi$.
For axion masses below $m_\Phi$, the cosmological constraints are more
restrictive, but the model is still acceptable for a wide range of
parameters.
We leave detailed investigation of the cosmology of these models for
future work.
The considerations above are sufficient to
show that the cosmology of this model is
acceptable for a wide range of parameters.

\section{CONCLUSIONS}
We have constructed a model in which the GUT scale is identified with 
the vacuum expectation value of a `GUT modulus' field with a potential that
is exactly flat in the supersymmetric limit.
In the context of gauge-mediated supersymmetry breaking, the potential
for the GUT modulus field arises at 2 loops, and pushes the modulus
to large values for a wide range of couplings.
If this runaway behavior is stabilized by Planck-suppressed operators,
the vacuum expectation value of the GUT modulus has the desired magnitude
independently of the scale of supersymmetry breaking,
as long as all dimensionless couplings are order one.
We find this mechanism to be quite compelling.
We have constructed an explicit model where this mechanism is embedded
in a realistic GUT with no dimensionful parameters and
natural doublet-triplet splitting.
We have also argued that the GUT modulus poses no
problems for cosmology provided there is an epoch of low-scale inflation.
This low-scale
inflation need not be responsible for structure formation.

This mechanism for fixing the GUT scale described here
is very simple, and there is no reason that it cannot
occur in other models.
For example, it should be possible to construct models based on $SO(10)$,
and there is no reason it cannot occur in a string GUT.

\section*{ACKNOWLEDGMENTS}
This work is supported by NSF
grant PHY-9421385,
and by a fellowship from the Alfred P. Sloan Foundation.


\end{document}